\documentclass[sigconf,9pt]{acmart}
\usepackage{booktabs} 
\hypersetup{bookmarks=false}
\usepackage{balance}
\renewcommand\footnotetextcopyrightpermission[1]{} 
\usepackage{latexsym}
\usepackage{amsmath}
\usepackage{dsfont}
\usepackage{amsfonts}
\usepackage{graphicx}
\usepackage{boxedminipage}
\usepackage{color} 

\usepackage{algorithm}
\usepackage{algorithmic}
\usepackage{subfigure}

\usepackage{bm}        

\bgroup\catcode`\#=12\egroup \newcommand{\eqdef}{\stackrel{\textrm{\tiny def}}{=}}
\usepackage{amsthm}
\usepackage[utf8]{inputenc}
\usepackage[english]{babel}
 
\setcopyright{rightsretained}
\acmDOI{10.475/123_4}
\acmISBN{123-4567-24-567/08/06}

\acmConference[IPSN'18]{ACM IPSN}{April 2018}{Porto, Portugal } 
\acmYear{2018}
\copyrightyear{2018}

\acmArticle{4}
\acmPrice{15.00}

\begin{document}
\title{Nanosecond-precision Time-of-Arrival Estimation for \\Aircraft Signals with low-cost SDR Receivers}

\author{Roberto Calvo-Palomino}
\affiliation{%
  \institution{IMDEA Networks Institute, Spain\\ University Carlos III of Madrid }
}
\email{roberto.calvo@imdea.org}

\author{Fabio Ricciato}
\affiliation{%
  \institution{University of Ljubljana, Slovenia}
}
\email{fabio.ricciato@fri.uni-lj.si}

\author{Blaz Repas}
\affiliation{%
  \institution{University of Ljubljana, Slovenia}
}
\email{br9404@student.uni-lj.si}

\author{Domenico Giustiniano}
\affiliation{%
  \institution{IMDEA Networks Institute, Spain}
}
\email{domenico.giustiniano@imdea.org}

\author{Vincent Lenders}
\affiliation{%
  \institution{armasuisse, Switzerland}
}
\email{vincent.lenders@armasuisse.ch}

\renewcommand{\shortauthors}{B. Trovato et al.}

\begin{abstract}
Precise Time-of-Arrival (TOA) estimations of aircraft and drone signals are important for a wide set of applications including aircraft/drone tracking, air traffic data verification, or self-localization. 
Our focus in this work is on TOA estimation methods that can run on low-cost  software-defined radio (SDR) receivers, as widely deployed in Mode S / ADS-B crowdsourced sensor networks such as the OpenSky Network. We evaluate experimentally classical TOA estimation methods which are based on a cross-correlation with a reconstructed message template and find that these methods are not optimal for such signals. We propose two alternative methods that provide superior results for real-world Mode S / ADS-B signals captured with low-cost SDR receivers. The best method achieves a standard deviation error of 1.5 ns.
\end{abstract}

\maketitle

\section{Introduction}

Aircraft and unmanned aerial vehicles continuously transmit wireless
signals for air traffic control and collision avoidance
purposes. These signals are either sent as responses to interrogations
by secondary surveillance radars (SSR) or automatically on a periodic
basis (ADS-B). Both types of signals are transmitted over the
so-called \emph{Mode S} data link~\cite{ModeS} on the 1090 MHz radio
frequency.

Over the last few years, sensor network projects have emerged which
collect those signals using a crowd of low-cost software-defined radio
(SDR) receivers such as e.g. the OpenSky Network~\cite{Schaefer14a},
Flightaware~\cite{flightaware}, Flightradar24~\cite{flightradar} and
many others.  These sensor networks can leverage the time-of-arrival
(TOA) of Mode S signals for various kinds of applications, including
aircraft localization~\cite{Schaefer14a,Strohmeier18}, air traffic
data verification~\cite{Schafer15,Strohmeier15,Strohmeier15b,Moser16,
  Jansen18}, and self-localization~\cite{Eichelberger17}.  In those
applications, a set of cooperating receivers measure \emph{locally}
the TOA of the arriving signals and then send these data to a central
computation server.  By joint processing the TOA of the same signal
arriving at different receivers, the central server is able to
estimate the location of the transmitter, the location of the
receivers, or the exact time when the signal was transmitted.

The accuracy of these applications heavily depends on the precision of
the TOA estimation, and in order to estimate positions up to a few
meters it is necessary to estimate the TOA with nanosecond
precision. The goal of this work is to provide a method for the TOA
estimation of Mode S signals that delivers nanosecond-level precision
even with low-cost SDR receivers, such as the widespread RTL-SDR
dongle \cite{silverspecs}.  We show that existing TOA estimation
approaches based on a cross-correlation with a reconstructed signal
template are sub-optimal in the particular context of Mode S
signals. In fact, the loose tolerance margins allowed by the
specifications on the shape and position of each individual symbol
within the packet (up to $\pm$ 50 ns) adds uncertainty to the
reconstruction of the whole packet waveform at the receiver.

We propose two alternative methods that improve the precision and at
the same time reduce the computational load.We test different variants
of TOA estimation on real-world signal traces captured with RTL-SDR,
which is currently the cheapest SDR device on the market and widely
used by crowdsourced sensor networks. Our results show that the best
proposed method delivers TOA estimates with a standard deviation error
of 1.5 ns. We further identify the limited dynamic range of the
RTL-SDR device (less than 50 dB with 8-bit analog-to-digital converter
(ADC) and fixed automatic-gain controller (AGC)) as the main
performance bottleneck, and show that sub-nanosecond precision is
achievable for signals that are not clipped due the limited dynamic
range of the device.

\section{Background}\label{sec:statement}

This section provides background on aircraft signals which we rely on
to estimate the TOA, and the limitations of classical TOA estimation
methods.

\subsection{Mode S signal format}
\begin{figure}[t]
\centering
 \includegraphics[width=0.9\columnwidth]{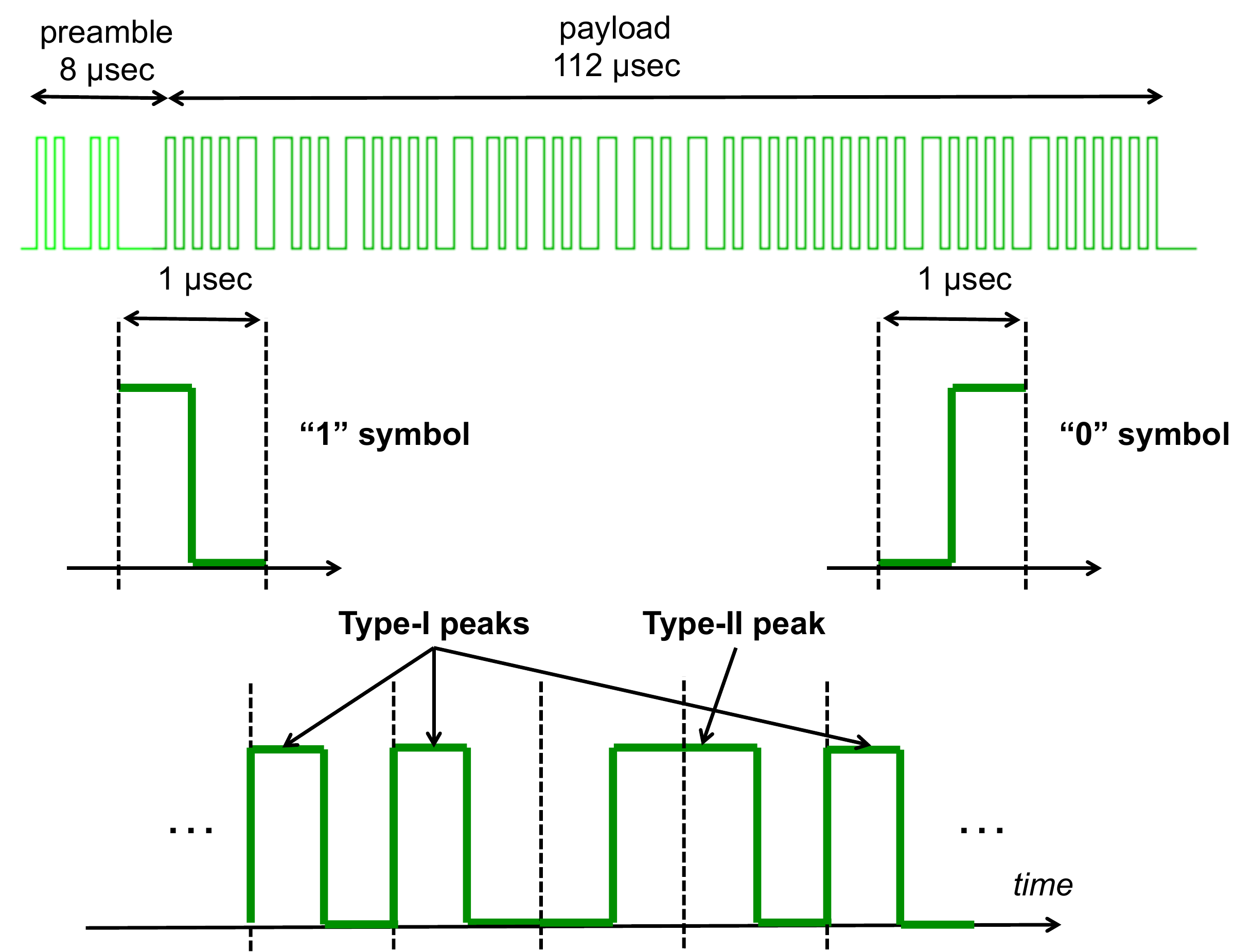}
\caption{Mode S packet structure with a binary PPM modulation.   }
 \label{fig:ppm}
\end{figure}

Hereafter we briefly review the physical-layer format of SSR Mode
S~\cite{ModeS2} reply and ADS-B messages transmitted by aircrafts on
the 1090 MHz channel.  Both packet formats consist of a preamble of 8
$\mu s$ plus a payload of 112 or 56 bits (only for other SSR Mode S
replies) sent at 1 Mbps rate, for a total duration of 120 $\mu s$ or
64 $\mu s$, respectively.  The information bits are modulated with a
simple Binary Pulse Position Modulation (BPPM) scheme as illustrated
in Fig. \ref{fig:ppm}: the symbol period of 1 $\mu$s is divided into
two ``chips" of 0.5 $\mu$s, and the high-to-low and low-to-high
transitions encode bits ``1" and ``0", respectively.  It is clear from
Fig. \ref{fig:ppm} that the BPPM modulation produces two types of
pulses of different duration, denoted hereafter as ``Type-I" and
``Type-II".  Type-I pulses have a nominal duration of one chip period
and are produced by the bit sequences ``00", ``11" and ``10". The
preamble consists of four Type-I pulses. On the other hand, Type-II
pulses have a nominal duration of two chip periods and are produced
exclusively by the ``01" sequence.

On average, we expect approximately $ 112/2=56$ Type-I and $ 112/4=28$
Type-II pulses for a payload of 112 bits.

The real-valued baseband signal is then modulated on the 1090 MHz
carrier frequency and transmitted over the air.  On the receiver side,
the decoding process relies exclusively on the signal
\emph{amplitude}, since in BPPM the signal \emph{phase} carries no
information.
\subsection{Limitation of standard TOA methods}
The standard ``course book" approach to TOA estimation in the Additive
White Gaussian Noise (AWGN) channel is a correlation filter
\cite{Kaybook}: the received signal is cross-correlated with a known
template corresponding to the source signal, and the point in time
maximizing the cross-correlation module is taken as TOA estimate.
   
The correlation method relies on the assumption that \emph{the source
  signal can be reconstructed very precisely} at the receiver, based
on the signal specifications and knowledge of the payload bits
$\bm{p}_m$.  Under this assumption, the correlation method represents
the Maximum Likelihood Estimator (MLE) \cite{Kaybook}. However, this
assumption is problematic in the particular case of \emph{real-world}
Mode S signals. In fact, the standard specifications tolerate up to
$\pm 50$ ns jitter in the \emph{position} of each individual pulse
within the packet: such high tolerance value is practically negligible
for the decoding process, but not for the task of determining the TOA
with nanosecond precision. As to the \emph{shape} of each pulse,
tolerance values of 50 ns are allowed for the pulse \emph{duration}
and \emph{rise time} and up to 150 ns for the \emph{decay time}, while
pulse amplitude may vary up to $2$ dB (approximately 60\%). Such loose
tolerance margins introduce uncertainty in the prediction of the shape
and position of the pulses in the source signal. Considering that Mode
S signals are typically received with high SNR, such an uncertainty
might well prevail over the effect of additive noise.  Consequently,
the correlation-based approach with a known packet template is no
longer guaranteed to be optimal, motivating the quest for alternative,
more precise methods.

\section{Our TOA estimation methods} 
In this section we describe the general approach to TOA estimation
based on the decoded payload and received signal samples, and then
present the different TOA estimation algorithms that were tested.
\label{sec:ourmethod}

\subsection{Signal acquisition architecture}
\begin{figure}[t]
\centering
\includegraphics[width=1\columnwidth]{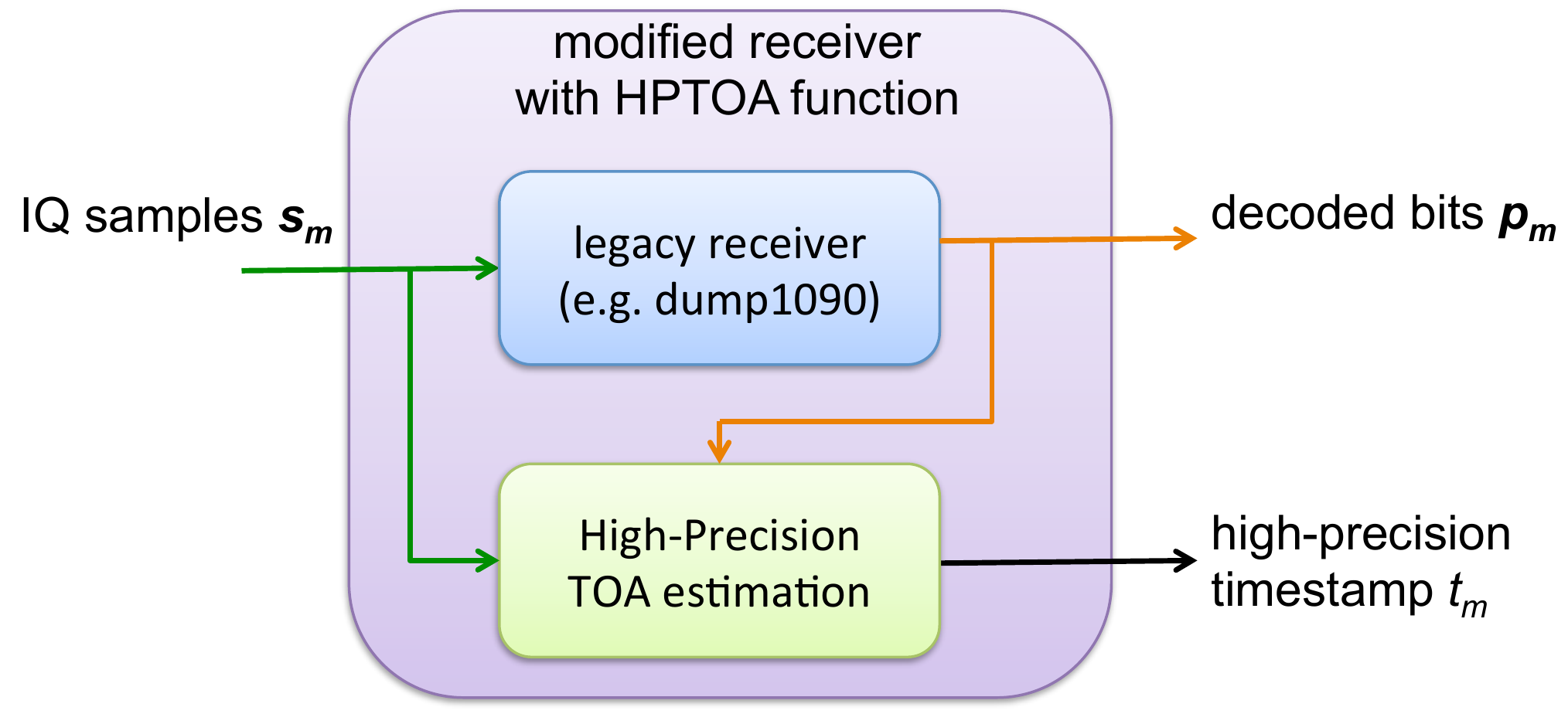}
\caption{Block diagram of improved receiver with high-precision TOA estimation. }
 \label{fig:legacyrx}
\end{figure}

In the software domain, the high-precision TOA estimation process can
be seen as an additional function that is optionally called within the
receiver and remains independent from the main decoding process. As
such, it can be implemented on top of any legacy receiver, including
but not limited to the widely adopted open-source tool
\texttt{dump1090} \cite{dump1090}. The overall block diagram of the
proposed scheme is exemplified in Fig.~\ref{fig:legacyrx}. The legacy
receiver takes as input a stream of complex in-phase and quadrature
(IQ) samples collected at sampling rate $f_s$ (for RTL-SDR hardware
$f_s=2.4$ MHz). The legacy receiver seeks to detect and decode the
incoming packet and, if successful, provides as output the decoded bit
sequence $\bm{p}_m$ along with the indication of the leading IQ sample
of the detected packet.
  
Denote by $\bm{s}_m$ the sequence of complex IQ samples corresponding
to the whole packet. The sequence includes approximately 300 samples
since we also pick a few samples immediately before and after the
packet in order to mitigate edge effects.  The sample vector
$\bm{s}_m$ and the decoded bit vector $\bm{p}_m$ represent the input
to our TOA estimation block.

\subsection{Proposed methods:  \emph{CorrPulse} and \emph{PeakPulse}}
\begin{figure}[t]
\centering
 \includegraphics[width=1\columnwidth]{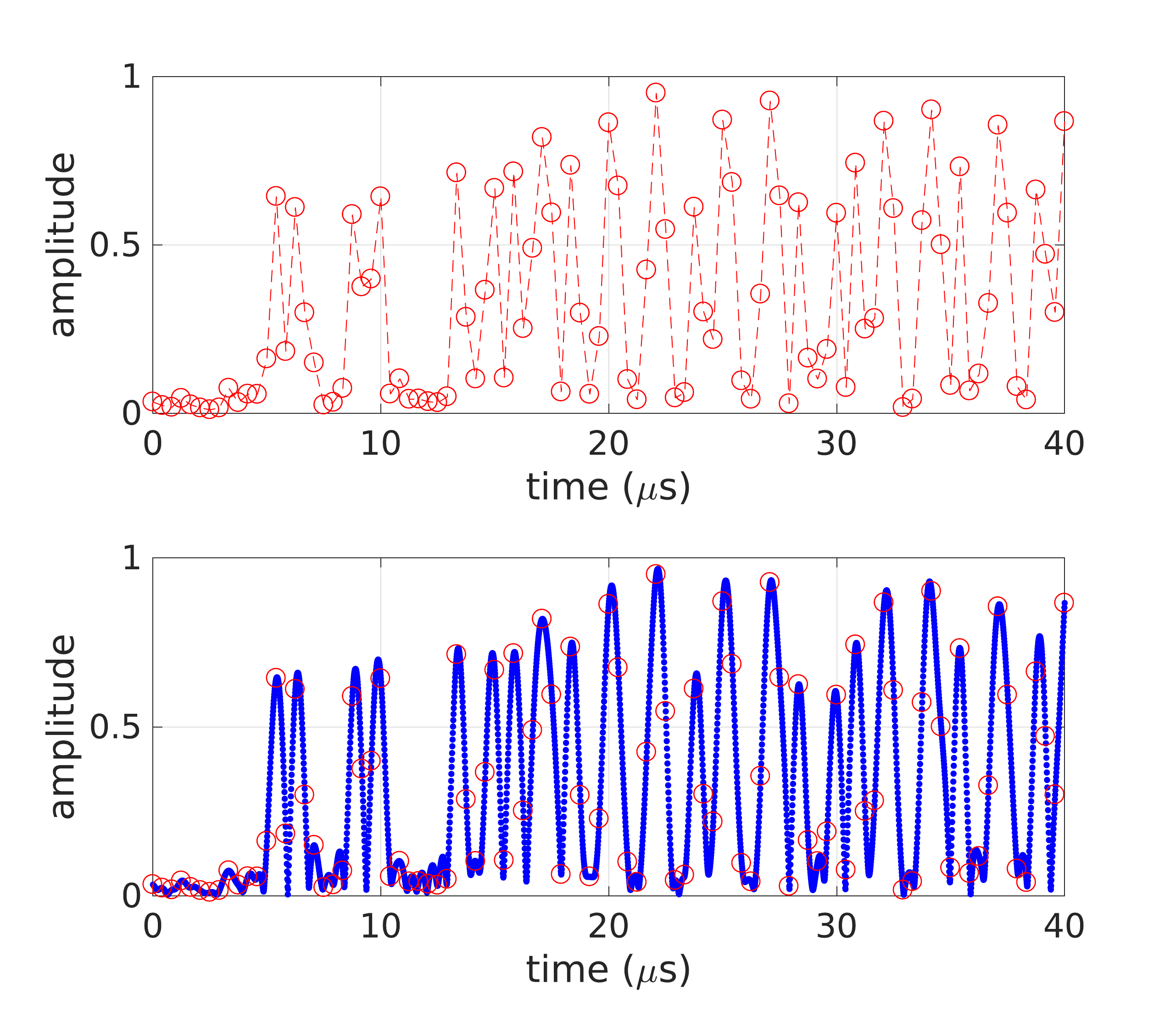}
\caption{Example of received signal amplitude corresponding to the
  preamble and initial payload of a real ADS-B packet. Original
  samples at $f_s=2.4$ MHz (top, red circles) and corresponding
  upsampled version (bottom, blue line).}
 \label{fig:upsampled}
\end{figure}

Hereafter we describe two novel TOA estimation algorithms specifically
developed for Mode S signals. For a generic packet $m$ we shall denote
by $K_m$ the total number of pulses in the whole packet (preamble and
payload). The input vector of complex samples $\bm{s}_m$ is
preliminarily upsampled by a factor $N$ and transformed into vector
$\bm{s}'_m$ (for a review of upsampling process see
e.g. \cite{Harrisbook}). To illustrate, Fig. \ref{fig:upsampled} plots
an excerpt of the amplitude of both vectors, namely $\left| \bm{s}_m
\right| $ (top plot) and $\left| \bm{s}'_m \right| $ (bottom plot),
for a generic packet found in a real-world trace.

The key aspect of the proposed algorithms is that the actual temporal
position $\hat{\tau}_k$ of the generic $k$th pulse within the packet
is estimated \emph{independently} from other pulses, with no need to
reconstruct a template for the whole packet. For each pulse $k\geq 2$,
we compute the individual shift $\Delta\tau_k \eqdef \hat{\tau}_k -
\tau_k$, i.e., the difference between the estimated and nominal pulse
position relative to the (estimated) position of the first
pulse. Finally, the pulse shifts are averaged in order to obtain the
final TOA estimate:
\begin{equation}
 \hat{t}  = \hat{\tau}_1 +  \frac{1}{K_m-1}\sum_{k=2}^{K_m}{\Delta\tau_k }
\end{equation}

The two proposed variants differ in the way individual pulse position
estimates are obtained, and which type(s) of pulses are considered.
In the first variant, labeled \emph{CorrPulse}, each pulse position is
determined through pulse-level cross-correlation of the upsampled
vector $\bm{s}'_m$ with the corresponding nominal pulse shape. Both
Type-I and Type-II pulses are considered in the final averaging.

In the second variant, labeled \emph{PeakPulse}, individual pulse
positions are determined by simply picking the local maximum point
value within the pulse interval, with no cross-correlation
operation. In this variant only Type-I pulses are considered, while
Type-II pulses are ignored. This is motivated by the fact that Type-II
pulses have lower curvature, hence their local peaks cannot be
identified as reliably as for Type-I pulses.

\section{Evaluation Methodology}

This section describes how we evaluate our new methods. First, we
introduce the other competing methods taken as reference for the
comparison. Then, we present the testbed setup with commercial
low-cost hardware. Finally, we provide details on the procedure
adopted to empirically assess the precision of the TOA measurement
methods in the given setup.

\subsection{Other methods for comparison}
\subsubsection{Correlation with whole-packet template: \emph{CorrPacket}}
This is the canonical cross-correlation method with a known signal template.

For every packet $m$, the whole packet template is reconstructed from
the decoded bits $\bm{p}_m$ and then cross-correlated with the
amplitude of the incoming signal. Here also, upsampling by a factor
$N$ is adopted to achieve sub-sample precision. Within the template,
the $k$th pulse is positioned at the nominal time $\tau_k$.  As to the
pulse shapes, we have tested two different variants: ``Rectangular"
(R), and ``Smoothed" (S). The two versions will be denoted by
\emph{CorrPacket/R} and \emph{CorrPacket/S}.  The rectangular pulses
have a nominal duration of 0.5 $\mu$s and 1 $\mu$s for Type-I and
Type-II pulses, respectively, and zero rise/decay times. The
rectangular pulse mask is represented exclusively by ``0" and ``1"
values, hence multiplications with another vector reduce to element
selection, which saves on computation load. The ``Smoothed" shape
corresponds to the output of a low-pass filter with passband of 2.4
MHz---matched to the bandwidth of the RTL-SDR receiver---when the
input signal is a nominal Type-I/Type-II pulse with the minimum
decay/rise time of 50 ns as per specifications \cite{modesspecs}.

\subsubsection{Existing dump1090 based implementations}
We also evaluate the precision of the timestamp reported by the
mutability fork of the open-source tool \texttt{dump1090}
\cite{dump1090}.  Furthermore, we test on our traces also the method
adopted by Eichelberger et al. in a recent ACM SenSys'17 paper
\cite{Eichelberger17} which is also based on dump1090. Code inspection
revealed that this method is based on a cross-correlation (implemented
in frequency domain) with a partial packet template consisting of the
preamble plus one quarter of the payload, with rectangular pulses and
upsampling factor $N=25$.
\subsection{Testbed setup}
\begin{figure}[t]
\centering
 \includegraphics[width=1\columnwidth]{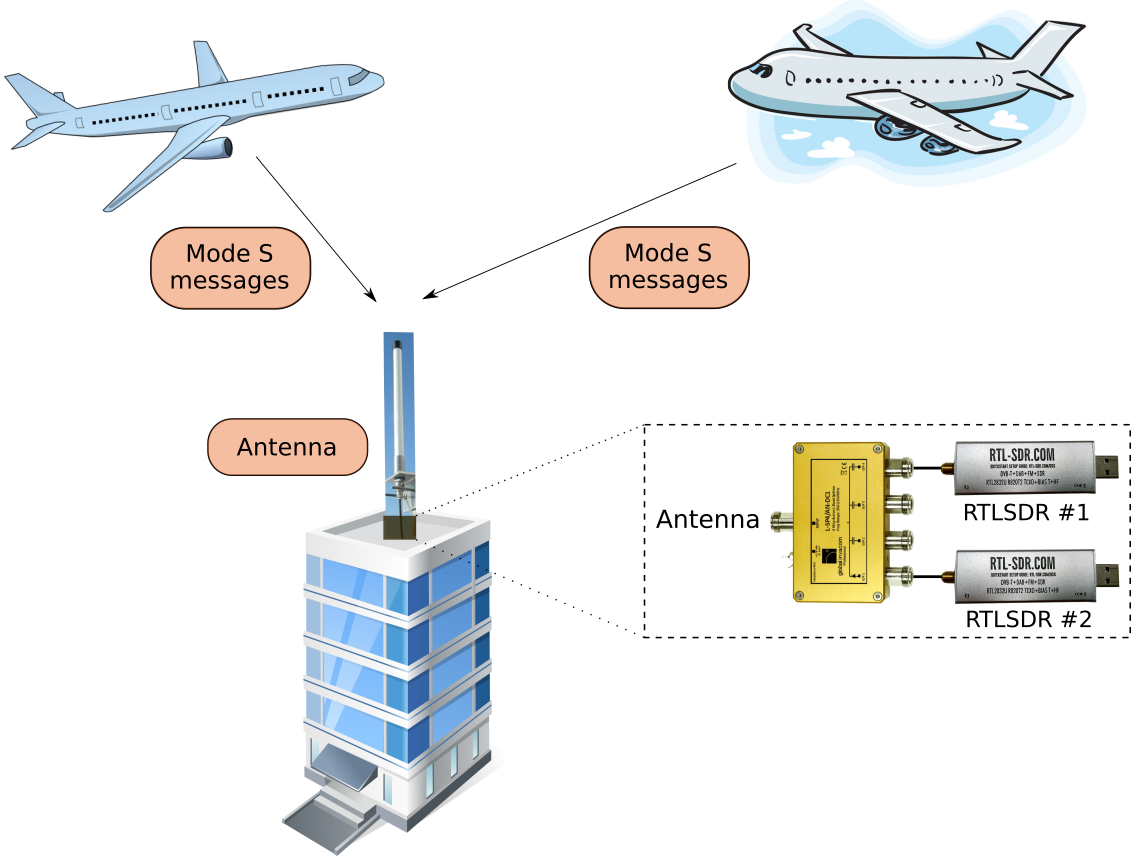}
\caption{Experimental setup. Two identical receivers connected to the
  same antenna via a splitter are collecting Mode S messages sent by
  aircraft.}
 \label{fig:testbed}
\end{figure}

The experimental setup consists of two identical sensors connected to
a single antenna through a power splitter and cables of identical
length. The sensors are located on the roof of a building as
Figure~\ref{fig:testbed} shows. Every sensor consists of one RTL-SDRv3
``Silver" model \cite{RTL-SDRv3} attached to a Raspberry Pi-3
\cite{rpi3}. The AGC gain is set to a fixed value, manually tuned to
maximize the packet decoding rate. The sampling rate was set to
$f_s=2.4$ MHz, the maximum value that our setup is able to acquire
with sample losses. Every I and Q sample is represented with 8
bit. The full stream of IQ samples are recorded one and processed
multiple times offline. Our results are based on a sample trace of 5
minutes collected in Thun (Switzerland) on 02-Aug-2017 at time 09:41.
The number of ADS-B packets that are correctly decoded \emph{at both
  sensors} by the \texttt{dump1090} open-source tool \cite{dump1090}
amount to 26445 from 59 different aircraft.

\subsection{Evaluation Metrics}\label{sec:clockerromodel}

In this section, we briefly describe the methodology adopted to assess
the precision of the different TOA estimation methods. The problem is
not trivial, since our receivers are not synchronized and the ``true"
TOA is unknown.  Therefore, we developed an evaluation method which
allows us to quantify the TOA precision without a ground truth. Denote
by $t_{m,i}$ the \emph{true} absolute arrival time of packet $m$ to
receiver $i$ and by $\hat{t}_{m,i}$ the corresponding \emph{measured}
TOA (by the method under test). In general, the measured value
$\hat{t}_{m,i}$ is affected by two distinct sources of error, namely
clock error and measurement noise:

\begin{equation}
\hat{t}_{m,i}= t_{m,i}  +  \xi_i(t)|_{t=t_{m,i}}+  \epsilon_{m,i}. 
\label{eq:clock1}
\end{equation} 
The term $ \xi_i(t)$ models the \emph{clock error} between the
receiver clock and the absolute time reference, and can be modeled by
a \emph{slowly-varying} function of time. Its magnitude depends on the
\emph{hardware} characteristics of the device, and specifically on the
stability of the local oscillator.

The term $\epsilon_{m,i}$ represents the measurement noise in the TOA
estimation process and is modeled by a \emph{random variable} with
zero mean and variance $\sigma^2_{\text{TOA}}$.  The \emph{precision}
of the TOA estimate, defined as the reciprocal of the noise variance,
is \emph{independent} of the clock error. The goal of the present
study is to reduce $\sigma^2_{\text{TOA}}$. The problem of
counteracting the clock error component remains outside the scope of
the present contribution. Here it suffices to mention that the clock
error can be mitigated by adopting receivers with GPS Disciplined
Oscillators (GPSDO), or it can be estimated and compensated in
post-processing \cite{ric18tmc,Galati12,gonzalo2005}.

\begin{figure}[t]
\centering
\subfigure[Low upsampling factor]{  \includegraphics[width=1\columnwidth]{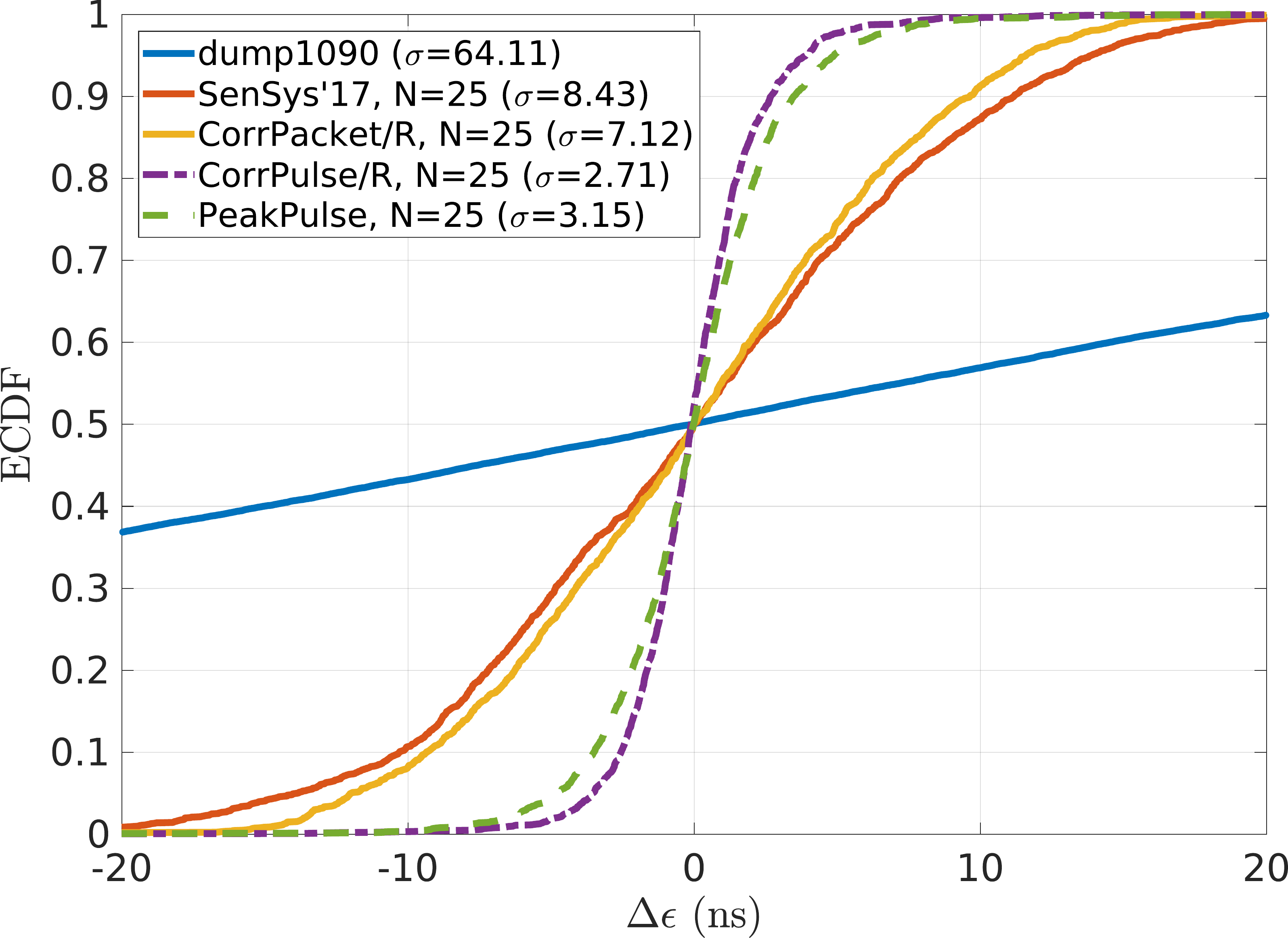} \label{fig_cdf25}} 
\subfigure[High upsampling factor]{  \includegraphics[width=1\columnwidth]{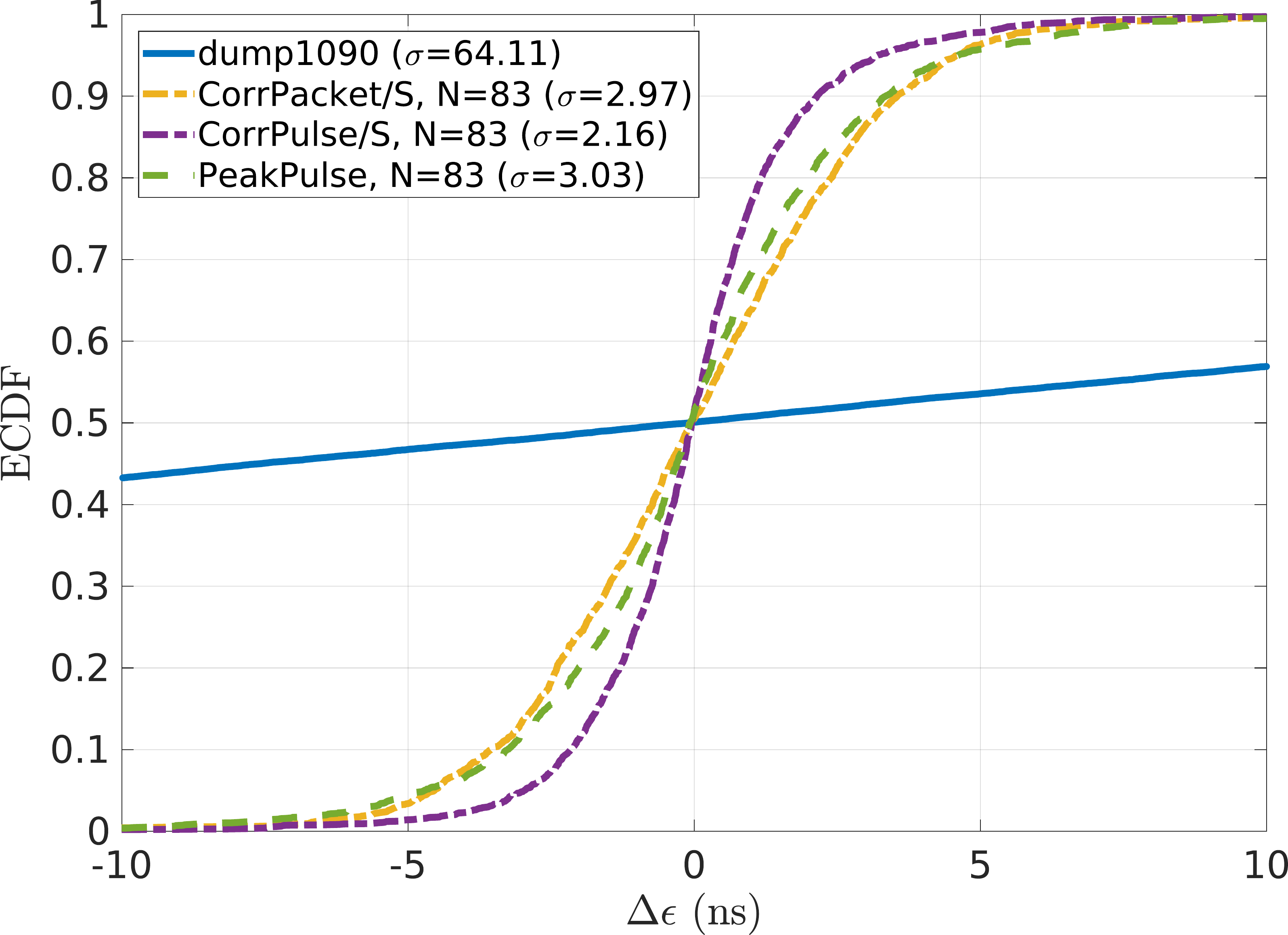} \label{fig_cdf83}}

\caption{ECDF of  $\Delta\epsilon$ residuals.}
 \label{fig:cdf}
\end{figure}

Hereafter we illustrate the methodology to experimentally quantify the
empirical TOA standard deviation $\hat{\sigma}_{\text{TOA}}$
notwithstanding the presence of a non-zero clock error component.
First, we need to get rid of the unknown \emph{true} absolute arrival
time $t_{m,i}$ in Equation \eqref{eq:clock1}. Since we use two
identical receivers attached to the same antenna, we can set
$t_{m,1}=t_{m,2}=t_m$ and subtract the TOA measurements at the two
sensors to obtain the corresponding time difference:
\begin{equation}
\hat{\Delta}t_m \eqdef \hat{t}_{m,2}-\hat{t}_{m,1}=  \Delta\xi(t_m) +  \Delta\epsilon_m
\label{eq:clock1bis}
\end{equation} 
wherein $ \Delta\xi(t) \eqdef \xi_2(t)-\xi_1(t)$ denotes the compound
clock error, and $\Delta\epsilon_m \eqdef
\epsilon_{m,2}-\epsilon_{m,1}$ the compound measurement error with
variance $\sigma^2_{\Delta\epsilon} =2 \sigma^2_{\text{TOA}}$. At
short time-scales, within the coherence time of the process
$\Delta\xi(t)$, the clock error represents a systematic error, i.e. a
\emph{bias} term that can be estimated and removed in order to
estimate the error \emph{variance} $\sigma^2_{\Delta\epsilon}$. We do
so by modeling the slowly-varying function $ \Delta\xi_i(t) $ by a
polynomial whose coefficients are estimated by standard
order-recursive Least Squares (refer to \cite[Chapter 8]{Kaybook} for
details).  After removing the estimated clock error component, we
obtain a set of residuals $\left\{\Delta\epsilon \right\}$. Their Mean
Square Error (MSE) represents an empirical estimate of twice the TOA
variance $MSE_{\Delta\epsilon}= 2\cdot
\sigma^2_{\text{TOA}}$. Accordingly, their Root Mean Square Error
(RMSE) provides a direct empirical estimate of the TOA error standard
deviation, formally:

\begin{equation*}\hat{\sigma}_{\text{TOA}} = \frac{1}{\sqrt{2}}\, RMSE_{\Delta\epsilon}  \approx 0.7 \cdot RMSE_{\Delta\epsilon}.
\end{equation*}

\section{Numerical Results}
We now present the results on the precision of the different TOA
estimation methods as evaluated in our testbed.
 \subsection{Error distribution}
In Fig. \ref{fig:cdf} we plot the Empirical Cumulative Distribution
Function (ECDF) of the residuals $\Delta\epsilon$'s obtained with
different TOA estimation methods for all the packets in the test
trace. The corresponding values of the TOA error standard deviation
$\hat{\sigma}_{\text{TOA}}$ are reported in the leftmost column of
Table \ref{table:results_tdoa}.

\begin{figure}[t]
\centering
 \includegraphics[width=1\columnwidth]{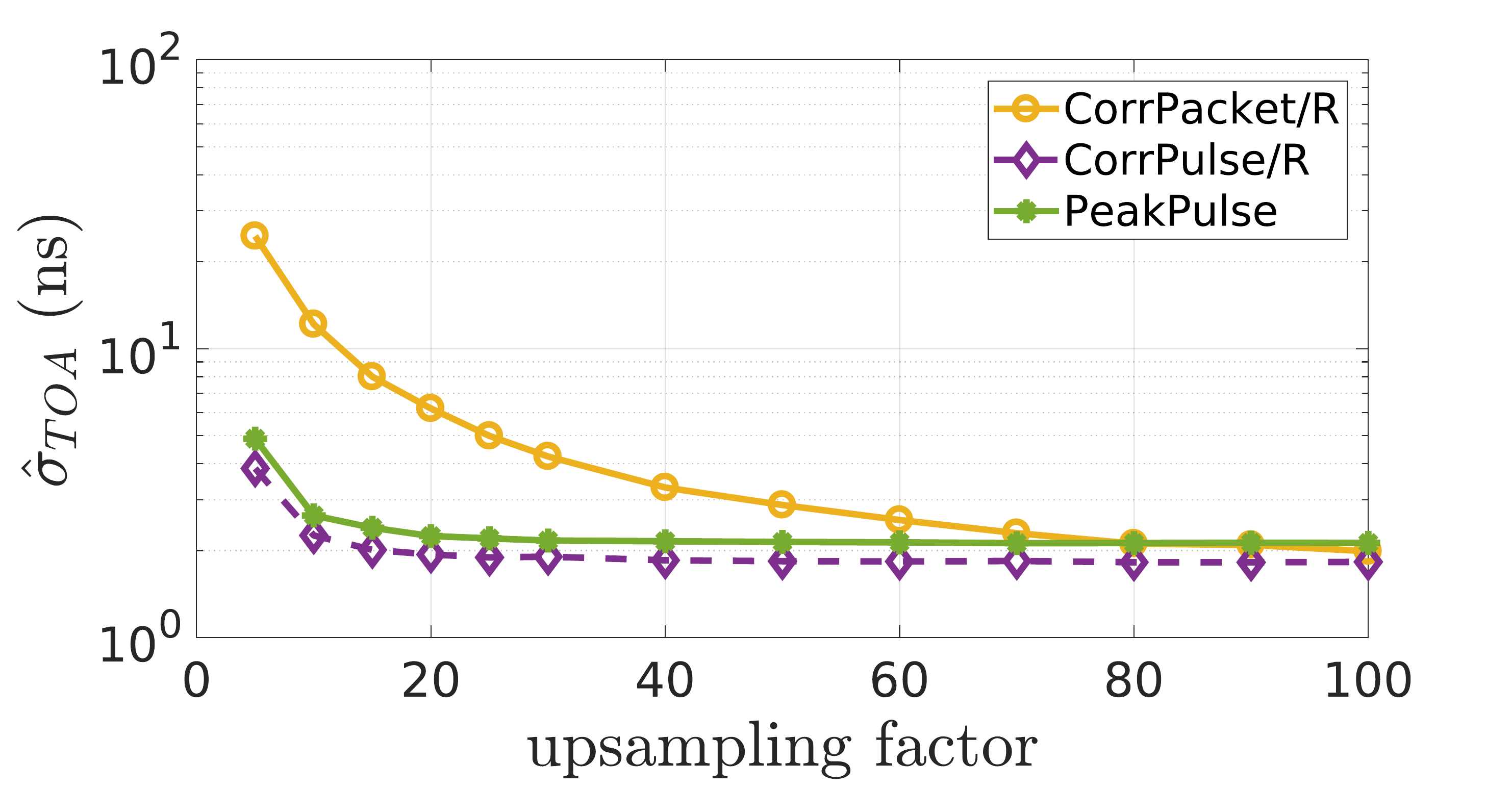}
\caption{TOA standard dev. error vs.  upsampling factor $N$}
 \label{fig:upsamplingfactors}
\end{figure}

For those applications where the computation load is of concern, it is
relevant to investigate the performance of the different methods with
moderate value of the upsampling factor ($N=25$). For
\emph{CorrPacket} and \emph{CorrPulse}, we consider the rectangular
pulse shape with binary 0/1 values, due to lower computation load.
Referring to Fig. \ref{fig_cdf25}, we observe that the proposed
\emph{PeakPulse} algorithm achieves a $RMSE_{\Delta\epsilon} = 3.15$
ns, less than half the value of the canonical \emph{CorrPacket/R}
method. It is remarkable that such good result was obtained with no
cross-correlation operation.  Fig. \ref{fig:upsamplingfactors} shows
$\hat{\sigma}_{\text{TOA}}$ for different values of the upsampling
factor $N$. We observe that the precision of the proposed methods
\emph{PeakPulse} and \emph{CorrPulse/R} improves faster than
\emph{CorrPacket/R} with increasing $N$.  These results indicate that
\emph{PeakPulse} should be preferred when computation load is at
premium.

Next we consider applications that enjoy abundant computation power,
for which the main goal is to maximize precision and computation load
is not of concern. For these, it is convenient to consider higher
upsampling factors ($N=83$ in our case) and, for the cross-correlation
methods, the more elaborated ``Smoothed" pulse shape. The latter
matches more closely the pulse shape passed through the RTL-SDR
front-end, leading to slightly higher precision than the simpler
``Rectangular" shape, as can be verified from Table
\ref{table:results_tdoa}. The ECDF of the residuals $\Delta\epsilon$'s
for these methods are plotted in Fig. \ref{fig_cdf83}. It can be seen
that the proposed \emph{CorrPulse/S} method is more precise than the
classical \emph{CorrPacket/S} method, and achieves
$RMSE_{\Delta\epsilon} = 2.16$ ns corresponding to
$\hat{\sigma}_{\text{TOA}}=1.51$ ns.

\subsection{Error vs. signal strength}
\begin{figure}[t]
\centering
 \includegraphics[width=.99\columnwidth]{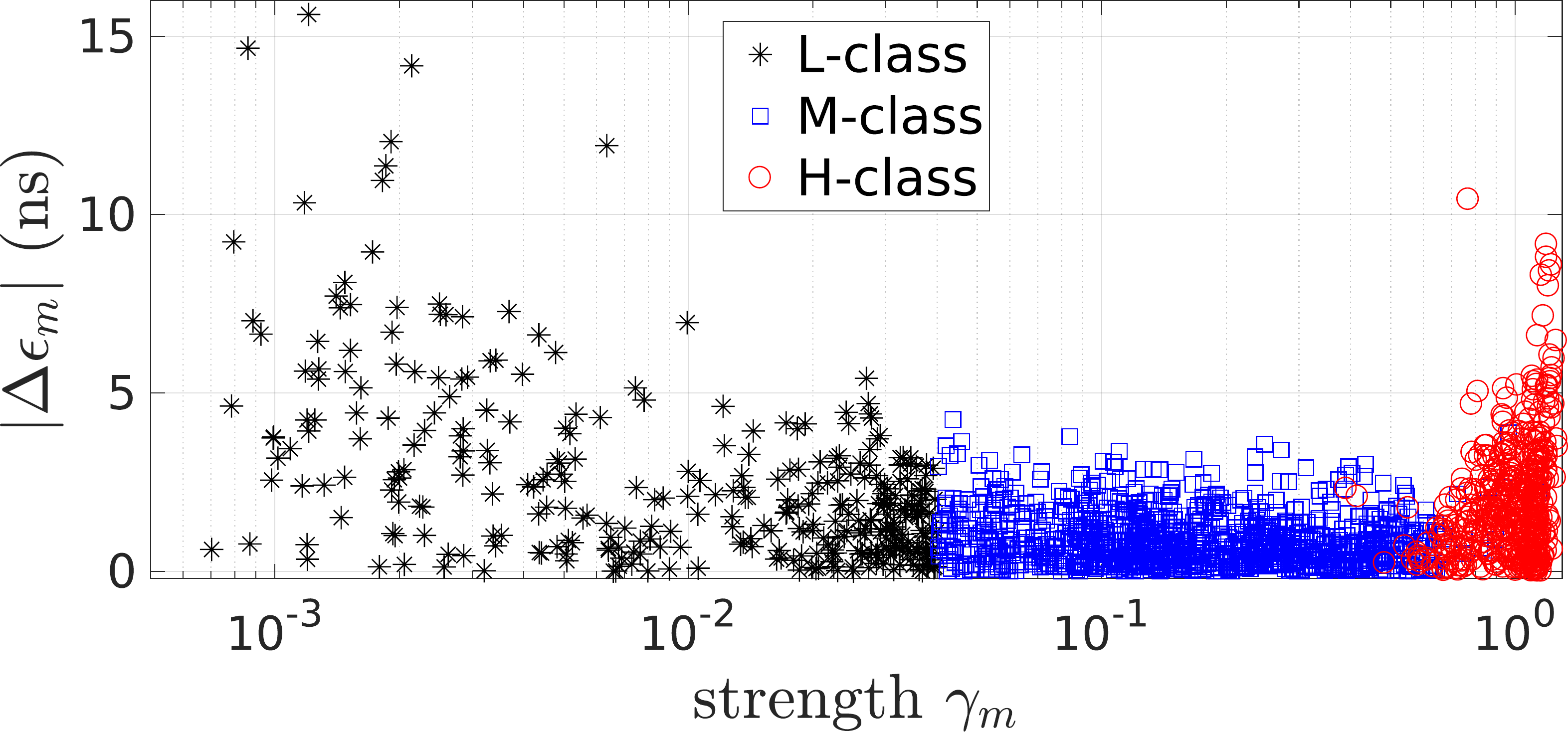}
\caption{Absolute error $\left| \Delta\epsilon_m\right|$ vs. packet strength $\gamma_m$.}
\label{fig:classes}
\end{figure}

In the following, we investigate the impact of signal strength on the
TOA error obtained with the most precise method, namely
\emph{CorrPulse/S with $N=83$}. For a generic packet $m$ and sensor
$i$, we denote by $\gamma_{m,i}$ the average of the squared pulse
height over all pulses --- an indicator of the arriving packet
strength. Furthermore, we denote by $\beta_{m,i}$ the number of pulses
that are clipped in the receiver due to one or more of the
corresponding IQ samples saturating the ADC. In
Fig. \ref{fig:classes}, we plot for each individual packet $m$ the
absolute value of the residual error $\left| \Delta\epsilon_m \right|
$ obtained with \emph{CorrPulse/S} ($N=83$) against the mean signal
strength between the two sensors $\gamma_m\eqdef
\frac{\gamma_{m,1}+\gamma_{m,2}}{2}$. Each packet is classified into
one of three classes: packets with $\gamma_m \leq 0.04$ are labeled by
``L", packets with $\min_{i =1,2}{\beta_{m,i}} \geq 10 $ are labeled
with ``H", and all remaining packets are labeled with ``M". The three
classes are marked respectively with black, red and blue markers in
Fig. \ref{fig:classes}. The estimated TOA error standard deviation
obtained by each method for each class are reported in Table
\ref{table:results_tdoa}. On one extreme, timing estimates for ``L"
packets with lower strength are impaired by quantization noise. On the
other extreme, packets received with high strength are subject to ADC
clipping, a form of distortion that clearly degrades timing precision.
As expected, these two classes yield higher error with all methods.
Between the two extremes, the strength of ``M" packets fits well the
dynamic range: for these, the proposed method achieves
$\hat{\sigma}_{\text{TOA}}=0.79$ ns.

\begin{table}[tb]\footnotesize 
\fontsize{10}{12}\selectfont 
\begin{center}
\begin{tabular}{ | c | c | c | c | c |}    
\hline
{\bf estimation method} & \multicolumn{4}{|c|}{\bf $\hat{\sigma}_{\text{TOA}}$ [nanoseconds]} \\
 &  all packets  & L  & M  & H  \\ 
 \hline
\hline 
legacy dump1090  &  $45.20$ & 44.94  & 45.19   & 45.43  \\
\hline
SenSys'17, $N=25$&  $5.90$ & 6.11 & 5.88 & 5.78 \\
\hline
\emph{CorrPacket/R}, $N=25$ &  $4.98$  & 5.48 & 4.85 & 4.94 \\
\emph{CorrPacket/R}, $N=83$ &  $2.14$  & 3.04 & 1.78  & 2.35  \\
\emph{CorrPacket/S}, $N=83$ &  $2.07$  & 3.00 & 1.68 & 2.275 \\ 
\hline
\emph{CorrPulse/R}, $N=25$ &  $1.89$  & 2.75 & 1.56 & 1.86 \\
\emph{CorrPulse/R}, $N=83$ &  $1.63$  & 2.72 & 1.04 & 1.77  \\
\emph{CorrPulse/S}, $N=83$ &  $1.51$  & 2.60 & 0.79 & 1.77 \\
\hline
\emph{PeakPulse}, $N=25$ &  $2.20$  & 3.36 & 1.70 & 2.23 \\
\emph{PeakPulse}, $N=83$ &  $2.12$  & 3.44 & 1.62 & 2.17 \\
\hline
\end{tabular}
\end{center}
\caption{Empirical estimates of TOA error standard deviation $\hat{\sigma}_{\text{TOA}}$.}
\label{table:results_tdoa}
\end{table}	

In our traces, less than 60\% of all packets fall into class
``M". With better hardware, and specifically with more ADC bits and
larger dynamic range, it would be possible to tune the AGC gain so as
to increase the fraction of packets falling in this class, thus
improving the overall precision.

The above results indicate that the received packet metrics
$\gamma_{m,1}$ and $\beta_{m,i}$ can be used to provide, \emph{for
  each individual TOA measurement} $\hat{t}_{m,i}$, also an indication
of the expected precision, i.e., of the error variance
$\hat{\sigma}^2_{m,i}$ affecting each individual measurement. In this
way, algorithms that take TOA measurements as input (e.g., for
position estimation) have the possibility to \emph{weight} optimally
each individual input measurement, as done e.g. in Weighted Least
Squares methods~\cite{strutz2010data}.

\begin{figure}[t]
\centering
\includegraphics[width=1\columnwidth]{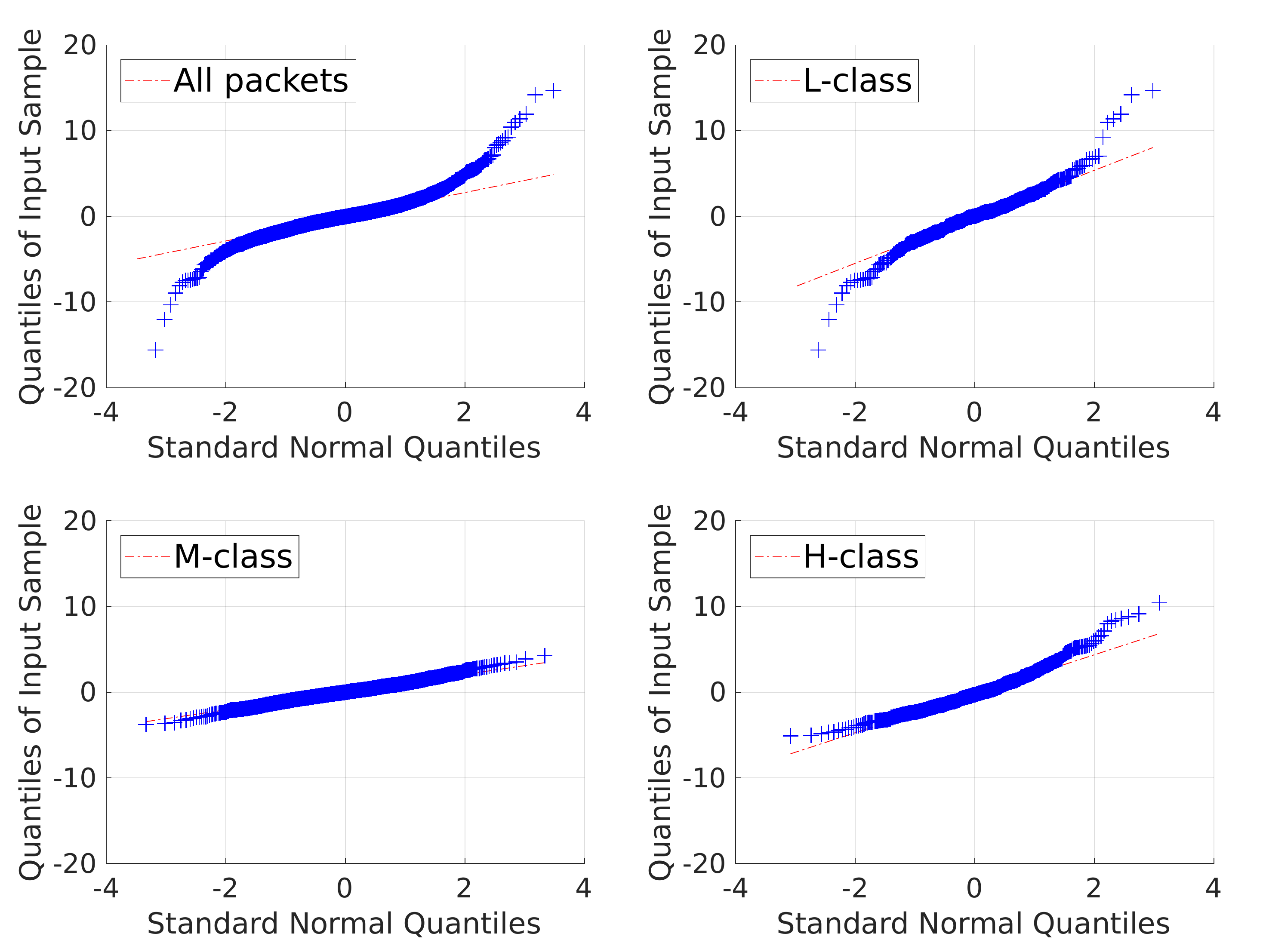}
\caption{Quantile-quantile plot of empirical errors $\Delta\epsilon$ vs. normal distribution.}
\label{fig:qqplot}

\end{figure}

Finally, we find that \emph{within each class} the empirical error
distribution is very well approximated by the Gaussian distribution,
as seen from the normal Q-Q plots in Fig. \ref{fig:qqplot}. This
justifies the adoption of Least Squares (LS) methods for position
estimation problems based on input TOA measurements \cite{ric18tmc},
since for normally distributed input errors the LS solution coincides
with the Maximum Likelihood estimate.

\balance
\section{Conclusions and Outlook}
We have presented two variants of a novel TOA estimation method for
Mode S signals that does not rely on long cross-correlations with the
template of a full packet. The most precise variant, namely
\emph{CorrPulse/S}, involves only short cross-correlation operations
on individual pulses. The other variant, namely \emph{PeakPulse}, is
lighter to compute, involves no cross-correlation operation and works
well also with moderate upsampling factors.

We have shown that such algorithms can achieve TOA estimates with nanosecond-level precision even with real-world signals captured with the cheapest SDR hardware that is currently available, namely RTL-SDR. A closer look at the test results reveals that the main limiting factor for the achievable TOA precision with RTL-SDR is the limited dynamic range --- less than 50 dB with 8-bit ADC and fixed AGC --- resulting in a large fraction of packets being  clipped  or  drowned into quantization noise. For packets that are received with signal strength well within the dynamic range of the receiver, the \emph{CorrPulse/S} achieves sub-nanosecond precision. It can be expected that  precision can be further improved with better hardware. The \emph{PeakPulse} method has been implemented in C, integrated in the dump1090 receiver and is released as open-source\footnote{http://github.com/openskynetwork/dump1090-hptoa}.

\section*{Acknowledgments}
This work has been funded in part by the Madrid Regional Government through the TIGRE5-CM program (S2013/ICE-2919). We would like to thank Manuel Eichelberger from ETH Zurich for sharing the code we used in our evaluation for comparison purposes. 

\bibliographystyle{ACM-Reference-Format}
\bibliography{toarob.bib} 

\end{document}